\documentstyle[sprocl]{article}

\def\be{\beta}
\def\ga{\gamma}
\def\de{\delta}
\def\ep{\epsilon}

\def\ka{\kappa}
\def\la{\lambda}

\def\si{\sigma}

\def\ph{\phi}

\def\ps{\psi}

\def\De{\Delta}

\def\cl{{\cal L}}

\def\frac#1#2{{\textstyle{{#1}\over {#2}}}}
\def\half{{\textstyle{1\over 2}}}
\def\lsim{\mathrel{\rlap{\lower4pt\hbox{\hskip1pt$\sim$}}
    \raise1pt\hbox{$<$}}}
\def\gsim{\mathrel{\rlap{\lower4pt\hbox{\hskip1pt$\sim$}}
    \raise1pt\hbox{$>$}}}
\def\sqr#1#2{{\vcenter{\vbox{\hrule height.#2pt
         \hbox{\vrule width.#2pt height#1pt \kern#1pt
         \vrule width.#2pt}
         \hrule height.#2pt}}}}

\def\lrprtmu{\stackrel{\leftrightarrow}{\partial_\mu}}
\def\lrprtnu{\stackrel{\leftrightarrow}{\partial^\nu}}

\def\Im{\hbox{Im}\,}

\newcommand{\beq}{\begin{equation}}
\newcommand{\eeq}{\end{equation}}
\newcommand{\bea}{\begin{eqnarray}}
\newcommand{\eea}{\end{eqnarray}}
\newcommand{\rf}[1]{(\ref{#1})}

\begin{document}

\begin{flushright}
IUHET 396\\
September 1998
\end{flushright}
\vglue 0.5 cm

\title{TESTING CPT AND LORENTZ SYMMETRY WITH\\
NEUTRAL-MESON OSCILLATIONS AND\\
QED EXPERIMENTS\footnote%
{Presented at the Workshop on CP Violation, 
Adelaide, Australia, July 1998} 
}

\author{V.\ ALAN KOSTELECK\'Y}

\address{Physics Department, Indiana University\\
Bloomington, IN 47405, U.S.A.\\
Email: kostelec@indiana.edu} 

\maketitle\abstracts{ 
This talk summarizes 
some theoretical features and experimental implications
of a general Lorentz-violating extension of the minimal 
SU(3) $\times$ SU(2)$ \times$ U(1) standard model
that allows for both CPT-even and CPT-odd effects. 
The theory would arise as the low-energy limit 
of a fundamental theory that is Lorentz and CPT covariant
but in which spontaneous Lorentz breaking occurs.
The use of neutral-meson oscillations 
and various QED systems
to bound the apparent CPT and Lorentz violations is described.
}

\section{Introduction}

A successful description of particle physics 
at presently attainable energy scales
is offered by the minimal standard model.
However,
at higher scales this is presumably replaced 
by an underlying theory incorporating
both gravity and quantum mechanics.
Experimental clues about the nature of the underlying theory
are difficult to obtain because the electroweak scale is
about 17 orders of magnitude smaller than the Planck scale
and so it is likely that any telltale effects
are heavily suppressed at present energies.

In seeking Planck-scale effects,
one method is to consider experimental searches
for physics that cannot occur 
in conventional renormalizable gauge theories.
Experiments of particular interest in this regard
are those with high sensitivity to
qualitatively new effects predicted in 
candidate underlying theories.
A promising example is string (M) theory,
for which such effects at the Planck scale
might generate low-energy signals. 

This talk considers the possibility 
that the new effects are generated from
spontaneous Lorentz symmetry breaking,\cite{kps}
which could arise in certain Lorentz-covariant theories
with suitable interactions 
among Lorentz-vector or tensor fields,
including perhaps some string theories.
If components of the tensor expectation values
generated by the spontaneous Lorentz breaking
are associated with the physical four spacetime dimensions,
then apparent violations of Lorentz symmetry could arise.
Since Lorentz invariance underlies 
the CPT theorem,\cite{sachs}
apparent breaking of CPT could also occur.

Any apparent violations of Lorentz and CPT symmetry
would be effects from the underlying theory
that are potentially observable
and that lie outside conventional renormalizable gauge theory.
In light of the probable heavy suppression,
detection of effects  
is likely only in highly sensitive experiments.

\section{Standard-Model and QED Extensions}

To include possible effects from  
spontaneous Lorentz and CPT violation 
in a description at the level of the standard model,
extra terms that break 
these symmetries and that are compatible with
an origin in spontaneous symmetry breaking
can be added to the lagrangian.
A general standard-model extension of this type,
including both CPT-even and CPT-odd terms,
has explicitly been given.\cite{kp,cksm}
It maintains the usual gauge structures,
including the gauge-symmetry breaking,
and is hermitian and power-counting renormalizable.
By construction,
it is the low-energy form of any underlying theory 
with spontaneous Lorentz and CPT violation
that generates the minimal standard model.

The origin of the standard-model extension
in a microscopic theory of spontaneous Lorentz violation
means that it exhibits many conventional properties
of Lorentz-covariant theories
despite the apparent Lorentz breaking.\cite{cksm}
Thus,
standard quantization methods can be applied,
and microcausality and positivity of the energy are expected. 
Also,
if the tensor expectation values 
arising from spontaneous symmetry breaking 
are independent of spacetime position,
energy and momentum are conserved.

Another attractive feature of the standard-model extension
is that the apparent noninvariance under Lorentz transformations
is restricted to
rotations or boosts of the (localized) fields only
(\it particle \rm Lorentz transformations).
Under rotations or boosts of the observer's inertial frame
(\it observer \rm Lorentz transformations),
the background tensor expectation values change along
with the field observables 
so the standard-model extension remains observer Lorentz 
covariant.\cite{cksm}
A related issue is the role of the  
Nambu-Goldstone modes that might be expected
from the spontaneous breaking of the global Lorentz symmetry
in the standard model.
The inclusion of gravity promotes Lorentz invariance
to a local symmetry,
and by analogy with the Higgs mechanism in gauge theories
one might expect these modes to generate a mass for the graviton.
It turns out, 
however,
that although the graviton propagator is modified
no gravitational Higgs effect occurs
because the gravitational analogue of the gauge field
is not the metric but involves instead 
its spacetime derivatives.\cite{kps}

Given the standard-model extension,
it is possible to extract a variety of interesting
limiting theories\cite{cksm}
that are relevant for experimental tests,
just as is normally done for the minimal standard model.
Among the useful theories that can be obtained 
are Lorentz-violating generalizations of the usual
forms of quantum electrodynamics (QED).
As an example,
in the special case 
of the theory of photons, electrons, and positrons,
the extended lagrangian includes extra terms in 
both the photon and the fermion sectors
that describe apparent Lorentz and CPT violations.
The usual QED lagrangian is 
\beq
\cl_{\rm QED} =
\overline{\ps} \ga^\mu (\half i \lrprtmu - q A_\mu ) \ps 
- m \overline{\ps} \ps 
- \frac 1 4 F_{\mu\nu}F^{\mu\nu}
\quad .
\label{a}
\eeq
There are several possible Lorentz-violating but
CPT preserving terms:
\bea
\cl^{\rm CPT~even}_{\rm QED} &=& 
c_{\mu\nu} \overline{\ps} \ga^{\mu} 
(\half i \lrprtnu - q A^\nu ) \ps 
+ d_{\mu\nu} \overline{\ps} \ga_5 \ga^\mu 
(\half i \lrprtnu - q A^\nu ) \ps
\nonumber\\ &&
- \half H_{\mu\nu} \overline{\ps} \si^{\mu\nu} \ps 
-\frac 1 4 (k_F)_{\ka\la\mu\nu} F^{\ka\la}F^{\mu\nu}
\quad ,
\label{c}
\eea
as well as Lorentz- and CPT-violating terms:
\beq
\cl^{\rm CPT~odd}_{\rm QED} =
- a_{\mu} \overline{\ps} \ga^{\mu} \ps 
- b_{\mu} \overline{\ps} \ga_5 \ga^{\mu} \ps 
+\half (k_{AF})^\ka \ep_{\ka\la\mu\nu} A^\la F^{\mu\nu}
\quad .
\label{b}
\eeq
Their coefficients are expected to be heavily suppressed
at accessible energy scales.
The reader can find more information about the above expressions
in the literature,\cite{cksm} 
including issues such as the observability 
of the new couplings.

\section{Experimental Tests}

The standard-model and QED extensions described above
provide a general microscopic theory of Lorentz and CPT violation
that can serve as a quantitative basis for experimental purposes. 
For example,
these include the identification 
of potentially sensitive experiments,
the analysis of data obtained,
and the comparison of bounds from different experiments. 

Relatively few experiments are sufficiently sensitive
to place constraints of interest on the extra coupling coefficients
in the standard-model extension.
Among those of exceptional sensitivity known to 
generate bounds on the standard-model and QED extension
are tests of CPT and Lorentz symmetry 
from
neutral-meson oscillations,\cite{kp,kexpt,ckpv,bexpt,ak}
Penning-trap measurements,\cite{pennexpts,bkr}
hydrogen and antihydrogen spectroscopy,\cite{antih,bkr2}
photon properties,\cite{cksm}
and baryogenesis.\cite{bckp}
The remainder of the talk briefly summarizes some of these results.
Studies presently being performed include
an analysis\cite{kla}
of constraints that could be obtained from 
clock-comparison experiments.\cite{cc}

\subsection{Neutral-Meson Systems}

There are four neutral-meson systems
in which oscillation experiments could be performed
to investigate CPT and Lorentz symmetry: 
$K$, $D$, $B_d$, and $B_s$.
In the following,
a neutral meson is generically denoted by $P$.

The effective hamiltonian for the time evolution
of a neutral-meson state is a two-by-two matrix with
complex entries.
Two kinds of (indirect) CP violation
can be studied within this framework.
One is the usual case of 
T violation with CPT invariance,
which for a $P$ meson is phenomenologically described 
by the standard parameter $\ep_P$.
The other is CPT violation with T invariance,
involving a complex parameter $\de_P$.

In the usual minimal standard model,
$\ep_P$ can be calculated in terms of other
parameters in the model
and $\de_P$ is identically zero.
However,
in the CPT- and Lorentz-violating standard-model extension
$\de_P$ is nonzero
and can be derived in terms of other parameters.\cite{ak}
It turns out that 
these include only the type of coupling coefficient 
that appears in certain CPT-breaking terms quadratic in the
quark fields $q$,
of the form $- a^q_{\mu} \overline{q} \ga^\mu q$.
In this expression,
the coupling $a^q_{\mu}$ is 
quark-flavor dependent but spacetime independent.
The derivation of the expression for $\de_P$ 
shows that it depends on $a^q_{\mu}$ 
as a direct result of flavor-changing effects.
Moreover,
experiments without flavor changes are insensitive 
to couplings of the type $a^q_{\mu}$,
so the results of CPT tests with neutral-meson oscillations
are independent of results of other CPT tests
such as those mentioned in the next subsection.

The conjuncture of Lorentz violations with the CPT breaking
also produces interesting effects.
For example, it can be shown that the parameter 
$\de_P$ 
depends on the boost and orientation of the meson.\cite{ak}
If the neutral-meson four-velocity is 
$\be^\mu \equiv \ga(1,\vec\be)$,
then at leading order in all coupling coefficients
$\de_P$ can be expressed as 
\beq
\de_P \approx i \sin\hat\ph \exp(i\hat\ph) 
\ga(\De a_0 - \vec \be \cdot \De \vec a) /\De m
\quad .
\label{e}
\eeq
In this equation,
$\De a_\mu \equiv a_\mu^{q_2} - a_\mu^{q_1}$,
where $q_1$ and $q_2$ denote the $P$-meson valence-quark flavors.
Also,
$\hat\ph\equiv \tan^{-1}(2\De m/\De\ga)$,
where $\De \ga$ and $\De m$ are the differences between the
decay rates and masses,
respectively,
of the $P$-meson eigenstates.
In Eq.\ \rf{e},
subscripts $P$ are understood on all quantities.

Equation \rf{e} shows that 
the size of $\de_P$ may differ for distinct $P$-meson flavors.
It is possible that relatively large CPT 
and Lorentz breaking occurs for non-kaon $P$ mesons,
for which few data are available.
Larger CPT violation in these systems would be plausible if,
for instance,
the couplings $a^q_\mu$ grow with the quark mass,
as occurs with the conventional Yukawa couplings.

Another result following from Eq.\ \rf{e}
is that the real part of $\de_P$ 
is proportional to the imaginary part.\cite{ckpv}
This equation also shows that $\de_P$ varies with boost magnitude
and orientation for a given $P$ meson,
which implies several interesting consequences.\cite{ak}
For instance,
experiments using mesons with large boosts
could be more sensitive to CPT violation
than otherwise comparable ones using mesons 
with lesser boosts because 
the CPT- and Lorentz-violating effects could be enhanced.

At present,
experiments on the kaon system have yielded the
tightest limits on CPT breaking.\cite{kexpt}
No bounds have as yet been extracted
from experiments with $D$ or $B_s$ mesons,
although for the $D$ and $B_d$ systems certain analyses 
of existing data can yield interesting constraints.\cite{ckpv} 
Indeed,
for the $B_d$ system,
two LEP collaborations\cite{bexpt}
at CERN have performed CPT studies with existing data.
A measurement of $\Im\de_{B_d}$ has been published by
the OPAL collaboration: 
$\Im\de_{B_d} = -0.020 \pm 0.016 \pm 0.006$.
A preliminary measurement has also been announced by 
the DELPHI collaboration:
$\Im\de_{B_d} = -0.011 \pm 0.017 \pm 0.005$.
Further investigations are being performed.

\subsection{Quantum Electrodynamics}

The remainder of this talk outlines some
of the implications of the standard-model extension
for a few QED experiments.

One important category of CPT and Lorentz tests
involves high-precision comparisons 
of particle and antiparticle properties.
A Penning trap can confine a single particle
for long periods during which
properties such as anomaly and cyclotron frequencies 
can be measured.\cite{pennexpts}
Predictions for signals in such experiments
have been extracted from the fermion sector
of the CPT- and Lorentz-violating 
standard-model and QED extensions.\cite{bkr}
Suitable figures of merit have been defined,
and estimates of the CPT and Lorentz reach
obtained.

As an explicit example,
measurements of the anomalous magnetic moments
of the electron and the positron 
are sensitive to
spatial components in the laboratory frame 
of the coupling $b_\mu$ in Eq.\ \rf{b}.
For these experiments,
a relatively minor change in experimental methodology
could produce\cite{bkr}
a bound on an appropriate figure of merit
of about $10^{-20}$.
Data from one such experiment are being analyzed.\cite{rm}
Another example is a comparison of
the cyclotron frequencies of antiprotons and hydrogen 
ions,\cite{gg}
for which the associated figure of merit could be
bounded\cite{bkr} at about $10^{-25}$.

Another class of important tests involves high-precision
spectroscopy of hydrogen and antihydrogen.\cite{ce}
An analysis of possible signals 
within the context of the 
standard-model and QED extensions
has been performed.\cite{bkr2}
It has been shown that
certain 1S-2S transitions and hyperfine Zeeman lines
are sensitive at leading order 
to Lorentz-violating effects.

The extra terms in the photon sector of the QED extension
are given in Eqs.\ \rf{c} and \rf{b}.
The CPT-even term in Eq.\ \rf{c} provides
a positive contribution to the energy\cite{cksm}
but the CPT-odd term in Eq.\ \rf{b} can generate
a negative one.\cite{cfj}
This is associated with some theoretical difficulties
that would seem to indicate the associated coupling
$(k_{AF})^\ka$ 
should vanish.\cite{cksm}

The extended Maxwell equations can be shown to 
describe the propagation of two independent
degrees of freedom with distinct dispersion relations.
However,
the Lorentz violation induces birefringence of the vacuum.
In the presence of the Lorentz violation,
an electromagnetic wave propagating in the vacuum
exhibits features closely related to those
found for the conventional Maxwell theory
when an electromagnetic wave travels 
in an optically anisotropic and gyrotropic transparent crystal 
exhibiting spatial dispersion of the axes.\cite{cksm}

Some experimental bounds can be placed
on the couplings in Eqs.\ \rf{c} and \rf{b}. 
An important limit is obtained from the
lack of observed anomalous birefringence 
of radio waves traveling over cosmological distances. 
The size of the components 
of the coupling $(k_{AF})_\mu$ 
are presently bounded to $\lsim 10^{-42}$ GeV,\cite{cfj,hpk}
although a disputed claim\cite{nr,misc}
has been advanced for an observed effect at the level of 
$|\vec k_{AF}|\sim 10^{-41}$ GeV.
The rotation-invariant irreducible component of 
the coupling $(k_F)_{\ka\la\mu\nu}$ 
is constrained to $\lsim 10^{-23}$ 
by several tests,
including the existence of cosmic rays.\cite{cg}
Other irreducible components
of $(k_F)_{\ka\la\mu\nu}$ 
violate rotation invariance.
A cosmological-birefringence bound of order $10^{-27}$ 
on the size of $(k_F)_{\ka\la\mu\nu}$ 
may be feasible with present methods.\cite{cksm}

\section*{Acknowledgments}
I thank Orfeu Bertolami, Robert Bluhm, Don Colladay, 
Rob Potting, Neil Russell, Stuart Samuel, 
and Rick Van Kooten for collaborations.
This work was supported in part
by the United States Department of Energy 
under grant number DE-FG02-91ER40661.

\end{document}